\documentclass[twocolumn,showpacs,preprintnumbers,amsmath,amssymb]{revtex4}

\usepackage{graphicx,subfigure}% Include figure files
\usepackage{dcolumn}% Align table columns on decimal point
\usepackage{bm}% bold math
\usepackage{gensymb}
\usepackage{wasysym}

\begin{document}

\preprint{APS/123-QED}

\title{Phase Field Modeling of Submonolayer Epitaxial Growth}% Force line breaks with \\

\author{Fan Ming}
\author{Andrew Zangwill}%
 \email{andrew.zangwill@physics.gatech.edu}
\affiliation{%
School of Physics \\ Georgia Institute of Technology \\ Atlanta, GA 30332\\
}%

\date{\today}

\begin{abstract}
We report simulations of submonolayer epitaxial growth using a continuum phase field model. The island density and the island size distribution both show scaling behavior. When the capillary length is small, the island size distribution is consistent with irreversible aggregation kinetics.  As the capillary length increases, the island size distribution reflects the effects of reversible aggregation. These results are in quantitative agreement with other simulation methods and with experiments. However, the scaling of the island total density does not agree with known results. The reasons are traced to the mechanisms of island nucleation and aggregation in the phase field model.
\end{abstract}

\maketitle

\section{Introduction}
Epitaxial growth is an important phenomena that has attracted theoretical attention from many different points of view.
The main motivation is to understand and predict the surface morphology as deposition proceeds. Some calculations focus on the energy parameters that control individual adatom motion. \cite{Mehl1999,Bogicevic1998} Other calculations focus on the kinetic roughening of the surface that occurs after thousands of layers have been deposited. \cite{Villain1998} The sub-monolayer regime is particularly interesting  because (i) comparison between experiment and theory can be used to extract  diffusion and adatom detachment barriers and (ii) the  kinetics of submonolayer growth is replicated in the  subsequent multilayer regime. \cite{Evans2006}

Several theoretical methods have been used to study the kinetics of sub-monolayer epitaxial growth. The oldest of these exploit rate equations to predict total island densities and the distribution of island sizes in a mean field theory. \cite{Sto,VENABLES1984} Kinetic Monte Carlo (KMC) simulations are particularly popular because they are atomistic, they provide a visualization of the growing surface, and they make predictions that often agree with experiment. \cite{BARTELT1992,RATSCH1995,Amar2002} A desire to avoid the computation-time restrictions of atomistic simulations led to the development of the continuum level set method (LSM), which focuses exclusively on the motion of steps. \cite{Chen1997,Gyure1998} Level set simulations have been shown to reproduce the results of KMC simulations for both sub-monolayer total island densities and island size distributions. \cite{Ratsch2000,Ratsch2001}

A recent paper by Yu and Liu \cite{Yu2004} approached the sub-monolayer problem using a phase field method. Phase field modeling is a continuum approach to the kinetics of phase transformations which makes no use of atomistic information. For that reason, it is widely used to study evolution phenomena over large length and time scales that are inaccessible to other methods. \cite{Emmerich} When applied to the problem of step flow growth in the limit of a thin interface (between the solid and its vapor), the phase field model reduces to the classic step flow model of Burton, Cabrera, and Frank. \cite{BURTON1951} Yu and Liu wrote down a phase field model to study the density of islands in the sub-monolayer regime. They reported that this quantity scaled with the deposition flux $F$ and the adatom surface diffusion constant $D$ as $N \propto (F/D)^{1/3}$. This is the expected result in the  irreversible aggregation regime where island nucleate when two atoms collide and there is no detachment of atoms from island edges.

The original motivation for this paper was to reproduce the island density results of Ref.~\onlinecite{Yu2004} and to extend them to study the distribution of island sizes in the sub-monolayer regime. It turned out that our results differed from theirs in a interesting way which, we believe, demonstrates some of the virtues and some of the defects of the phase field method applied to this particular problem. Our main result is that the island size distribution shows scaling behavior. When the capillary length is small, the island size distribution is consistent with irreversible aggregation kinetics.  As the capillary length increases, the islands size distribution reflects the effects of reversible aggregation. The results agree quantitatively with KMC and LSM simulations and with experimental data. The total island density scales with $D/F$, but the exponent is not ${1\over 3}$, nor does it change when the scaled island size distribution changes shape.

\section{Calculational Method}

The phase field model of Yu and Liu uses two dimensionless variables, the adatom concentration $u$ and the order parameter (surface profile)  $\phi$. These are coupled by the evolution equations:
\begin{eqnarray}
\frac{\partial u}{\partial t} &=& D\nabla^2 u - \frac{\partial \phi}{\partial t} + F +\eta \label{density}  \\ & & \nonumber  \\
\frac{\partial\phi}{\partial t} &=& \frac{1}{\tau}\{W^2\nabla^2\phi - 2 \sin (2\pi\phi) \nonumber \\ & & \nonumber \\
 & -& \lambda (u-u_{eq})[2\cos (2\pi\phi) - 2] \} + \lambda_n D u^2.   \label{phase}
\end{eqnarray}
In (\ref{density}), the first term models the surface diffusion of adatoms.  The second term models mass exchange between the adatom population and the steps. The third term is the mean deposition rate, and the last term is a random variable which determines the points on the surface where deposited atoms land. In (\ref{phase}), the term $2\sin(2\pi \phi)$ identifies the terraces of the step profile with integer values of  $\phi$. The term $W^2\nabla^2\phi$ determines the width $W$ of the step which  connects adjacent terraces and the term proportional to $u-u_{eq}$ causes the boundary of an island to move by the capture or release of adatoms. The final term in (\ref{phase}) is a rate equation estimate of the island nucleation rate.

To discuss our choice of  parameters, we recall the ``thin-interface'' limit of the phase field model. \cite{Karma1996} This limit defines a capillary length and a kinetic coefficient $\beta$ from
\begin{equation}
\label{d0}
d_0 = a_1\frac{W}{\lambda},
\end{equation}
and
\begin{equation}
\label{beta}
 \beta = \frac{a_1 \tau}{\lambda W}\left[ 1-a_2\lambda\frac{W^2}{D\tau} \right],
 \end{equation}
 where $a_1=0.36$ and $a_2=0.51$. More importantly, $d_0$ and $\beta$ are related to each other in exactly
 the same way as they are related in the Burton, Cabrera, and Frank model of step flow growth. \cite{Pierre-Louis2003} Namely,
 \begin{equation}
 \label{u-step}
 v=D[\hat{\bf n}\cdot \nabla u]_{step}=\beta^{-1}[u-u_{eq}-d_0 \kappa]_{step}
% \beta v = u|_{step}-(u_{eq}+d_0\kappa),
 \end{equation}
 where $v$ is the velocity of a step, $\hat{\bf n}$ is a unit vector normal to the step,  $u_{eq}$ is the equilibrium concentration of adatoms at a straight step,
 and $\kappa$ is the step curvature. The subscript ``step'' in (\ref{u-step}) means that the quantities
 in brackets are evaluated at the step edge.
We consider the limit $\beta=0$ only, which corresponds to fast attachment of adatoms to step edges  (surface diffusion limited growth). In that case, we get the Gibbs-Thomson equation \cite{Villain1998}
\begin{equation}
[u]_{step} = u_{eq}+d_0[\kappa]_{step},
 \label{uu-step}
\end{equation}
and there is no loss of generality if we set $u_{eq}=0$. In
the same $\beta=0$ limit,
\begin{equation}
\lambda= \frac{a_1 W}{d_0}~~~~~~~~~~{\rm and}~~~~~~~~~\tau = \frac{a_1 a_2 W^3}{d_0 D}.
\end{equation}

In practice, we chose a unit length $a$ and fixed $W=a$ and $D=10^4 a^2/{\rm sec}$. The free parameters of the model are  $d_0$ (units of $a$), $F$ (units of ML/{\rm sec}), and $\lambda_n$. We discretized the coupled equations (\ref{density}) and (\ref{phase}) on a $L \times L$ square lattice with $L=960$ grid points and solved them using no-flux boundary condition at the lattice edges and  a two-dimensional forward-time, central space (finite-difference) algorithm. A parallel algorithm (domain decomposition) was used to speed up the computation. We found good convergence using a spatial grid size $\Delta x=0.4a$. The time step $\Delta t$ is chosen so that $\Delta t \ll (\Delta x)^2/D$. To model depositions, we choose a grid site at random and set $u=a^2/(\Delta x)^2$ at that site. We then repeated this step every $1/(F L\Delta x)^2$ seconds. The surface coverage is defined as $\theta=Ft$.

\section{Results}
\subsection{Nucleation \& Aggregation}
\begin{figure}
\includegraphics[trim = 20.5mm 35mm 2mm 33mm, clip, scale=0.5]{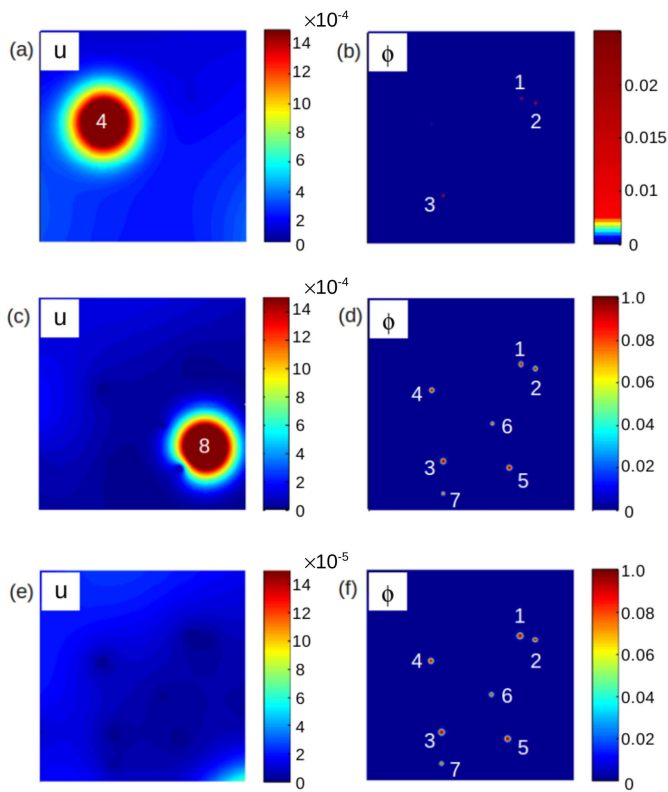}
\caption{Time evolution of the order parameter and corresponding adatom concentration.
$\theta$ is the surface coverage. Note that the color bar is varied to optimize the contrast. For all panels, $D/F=10^7$,
$d_0=1.44\times 10^{-6}$, $\lambda_n=8.4\times 10^{-3}$ and $L=80a$. The surface coverage: panel (a) and (b) $\theta=2.7\times10^{-4}$, panel (c) and (d): $\theta=2.2\times10^{-2}$, panel (e) and (f): $\theta=2.8\times10^{-2}$}.
\label{fig:side}
\end{figure}

Figure~\ref{fig:side} illustrates the nucleation and aggregation behavior produced by the phase field equations (\ref{density}) and (\ref{phase}). The left column shows the adatom density $u$ at three successive times. The right column shows the order parameter $\phi$ (surface morphology) at the same three times. Panel~(a) shows the rapid, isotropic diffusion of the adatom concentration away from a deposition event which occurred at the point labelled (4). Through the nucleation term in (\ref{phase}), this distribution of $u$ triggers the growth of a small spike in $\phi$ at exactly the point (4). This spike, which we call a proto-island, is not yet visible in panel~(b), which instead shows three proto-islands [labelled (1)-(3)] which were triggered by three earlier deposition events. The adatom density associated with these earlier events has completely diffused away by the time of deposition event (4).

Understanding the fate of proto-islands is the key to understanding the behavior of the model overall. Some proto-islands grow into true islands by the capture of adatom density from other deposition events. Other proto-islands disappear because not enough adatom density is captured before $\phi$ itself ``diffuses'' away due to the interface width term $W$ in (\ref{phase}). Our choice of $W$ produces well-defined islands with sharp edges. Diffusion along the island edges is naturally included by the surface free energy minimization that leads to (\ref{phase}). In detail, we label as a proto-island  every set of one or more nearest-neighbor connected grid sites where $\phi>0.05$. If the value of $\phi$ at each connected site is called $\phi_k$, we form the quantity $s=(\Delta x/a)^2\sum_k\phi_k$ for each proto-island and monitor its value as time goes on. If $s \to 0$, we say that this proto-island has disappeared; if $s>1$ we say this proto-island has become a true island composed of $s$ atoms.

Panel~(c) in Figure~\ref{fig:side} shows the expected adatom concentration very soon after a deposition event at the point labelled (8). More interesting is
panel~(d), which shows seven true islands. Islands (1)-(3) evolved from the  proto-islands (1)-(3) in panel~(b). Islands (4)-(7) were produced by deposition events that occurred in the time between panels~(a) and (c). A short time later, panel~(e) shows that the adatom density associated with deposition event (8) has diffused entirely away. However, no island (8) has been created in panel~(f) because proto-island (8) disappeared. It did not grow to a true island because the existing islands captured all the available adatom density. In other words, the island density in this neighborhood of the surface has saturated and further deposition only causes the existing islands to grow. Indeed, the very dark regions of panel~(e) can be regarded as ``denuded'' zones around each island.

The foregoing shows that the {\it nucleation} of an island in the phase field model occurs quite differently than it does in, say, an atomistic KMC simulation. There, deposited atoms diffuse on the surface until they collide to form a stable island somewhere away from the deposition point of either atom. We have said that the phrase  ``irreversible growth'' is used if this collision produces a stable island. We speak of  ``reversible growth'' if a just-nucleated island can dissociate back into adatoms. That being said, the {\it aggregation} behavior of the phase field model seems quite similar to that seen in KMC and LSM simulations. We will see in a moment that this similarity (dissimilarity) of the nucleation (aggregation) process to other simulation results has consequences for the behavior of the distribution of island sizes and for the total island density.

For later use, we draws particular attention to the level set method to simulate sub-monolayer epitaxial growth.  In LSM simulations,  islands are nucleated at random positions on the surface using a rate-equation-like  weighting factor proportional to the square of the adatom density. \cite{Ratsch2000} The adatom density itself evolves as dictated by a uniform deposition flux at every point  and a diffusion equation with specified boundary conditions at the moving edges of existing islands. The method is very computer-time intensive, but as mentioned earlier, the total island density and the distribution of island sizes agree very well with KMC simulations and with experiment.

\subsection{Island Size Distribution}

The island size distribution $n_s$ is the number of islands composed of $s$ atoms. If $s_{av}$ is the average island size, it is well-known that a plot of the scaled quantity $n_s s^2_{av}/\theta$ versus $s/s_{av}$ will collapse onto a single curve data collected for different values of $D/F$. \cite{BARTELT1992,Evans2006} One particular curve is characteristic of irreversible aggregation and the shape of this curve varies smoothly as the degree of reversibility is increased by changing, say, the pair-bond energy in a KMC simulation. \cite{RATSCH1995}

\begin{figure}
\includegraphics[trim = 80mm 0.5mm 30mm 5mm, clip, scale=0.4]{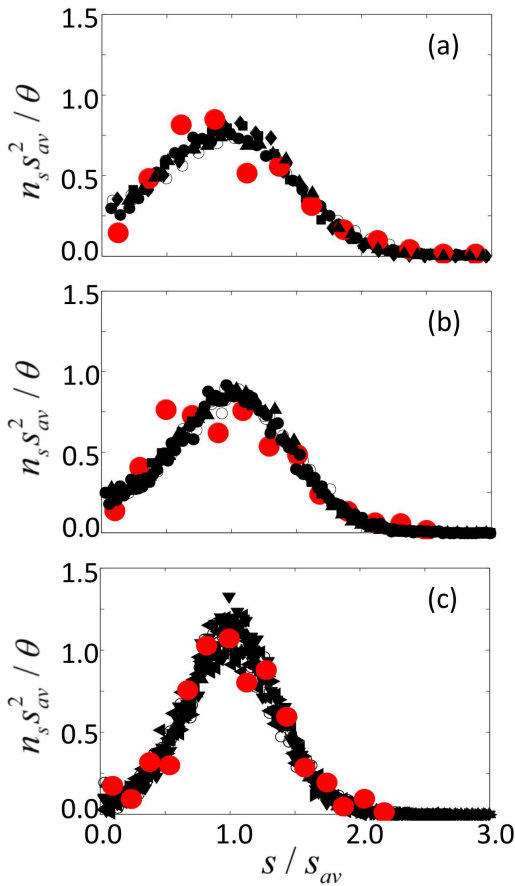}
\caption{The crossover scaling of island size distribution. Experimental data (large circles) are replotted from \cite{STROSCIO1994} for different temperatures, and KMC data (open symbols) from \cite{RATSCH1995}. (a) $\blacksquare$: $D/F=10^5, d_0=1.44\times10^{-4}, \theta=0.06, \lambda_n=0.03$; \CIRCLE \ and $\blacktriangle$: $D/F=10^6, d_0=1.44\times10^{-5}$ and $2.43\times10^{-5}, \lambda_n=0.06$ and $0.1, \theta=0.05 - 0.1$. $\blacklozenge$: $D/F=10^7, d_0=1.44\times10^{-6}, \lambda_n=8.4\times10^{-3}, \theta=0.01$. (b) $\blacksquare$: $D/F=10^5, d_0=1.44\times10^{-4}, \theta=0.1, \lambda_n=0.03$; \CIRCLE \ and $\blacktriangle$: $D/F=10^6, d_0=1.44\times10^{-5}$ and $4.0\times10^{-5}, \lambda_n=0.012$ and $0.1, \theta=0.05 - 0.1$. (c) $D/F=10^6$. $\blacktriangle$ and $\blacktriangledown$: $d_0=1.0\times10^{-4}, \lambda_n=0.1$, $\theta=0.05$ and $0.1$; $\blacktriangleright$ and $\blacktriangleleft$: $d_0=3.2\times10^{-4}, \lambda_n=1$, $\theta=0.05$ and $0.1$. }
\label{fig:distr}
\end{figure}

Fig.~\ref{fig:distr}(a) shows island size distributions obtained from our phase field simulations model at very low coverage for $D/F=10^5-10^7$ and various choices of the model parameters $d_0$ and $\lambda_n$.  Each data point of the same symbol represents the average of at least 20 simulations. The scaling curve we find agrees very well with {\it irreversible} KMC and LSM simulations and with low temperature experimental data collected for Fe/Fe(001). \cite{RATSCH1995, STROSCIO1994}  Data collapse onto a single  curve generally required us to reduce the value of $d_0$ as we increased the value of $D/F$.  Doing this (or changing $\lambda_n$)  produced very different total island densities,  even though the scaled island size distributions were the same. For example, the data associated with the symbols $\blacktriangle$ and \CIRCLE \ in Fig.~\ref{fig:distr}(a) have island densities that differ by $25\%$. Similar behavior occurs in LSM simulations when the boundary conditions at the island edges are changed slightly. \cite{Ratsch2001} Based on Figure~\ref{fig:distr}(a), we conclude that the  details of the island nucleation process are not critical to the  shape of the island distribution when irreversible growth occurs. What matters is the subsequent process of monolayer capture by existing islands.

Fig.~\ref{fig:distr}(b) and Fig.~\ref{fig:distr}(c) show the effect on the island size distribution of progressively increasing the capillary constant $d_0$. The $\blacktriangle$ data in these two figures correspond to the same choices of $D/F$, $\lambda_n$, and $\theta$ used in Fig.~\ref{fig:distr}(a). The change in shape we find for the scaled island size distribution as $d_0$ increases agrees quantitatively with the change in shape seen in  {\it reversible} KMC simulations when the pair-bond energy is decreased or (equivalently) when the critical island size is increased.  \cite{Amar2002} Our results also agree with {\it reversible} LSM simulations. \cite{Petersen2001}

The step velocity in reversible LSM simulations is calculated from
\begin{equation}
v = D[\bm{n}\cdot\nabla u]_{step} - v_{det}, \label{detachment}
\end{equation}
where the second term  takes account of the detachment of atoms from island boundaries. Typically,  $v_{det}$ is taken to be proportional to the density of island  edge atoms. This may be contrasted with our (\ref{uu-step}), which  shows that increasing $d_0$ has the effect of raising the adatom density at islands edges (which is zero in LSM simulations). For the BCF problem of adatom diffusion on terraces, this simultaneously reduces the gradient of the adatom density at the step edge in the leftmost equation in (\ref{u-step}) and thus retards the growth speed of an island.
The capillary constant $d_0$ measures the strength of the Gibbs-Thomson effect, \cite{Villain1998} which is the driving force for adatom detachment from step edges in phase field modeling.

\subsection{Total Island Density}
\begin{figure}
\includegraphics[trim = 15mm 0.5mm 15mm 4mm, clip, scale=0.35]{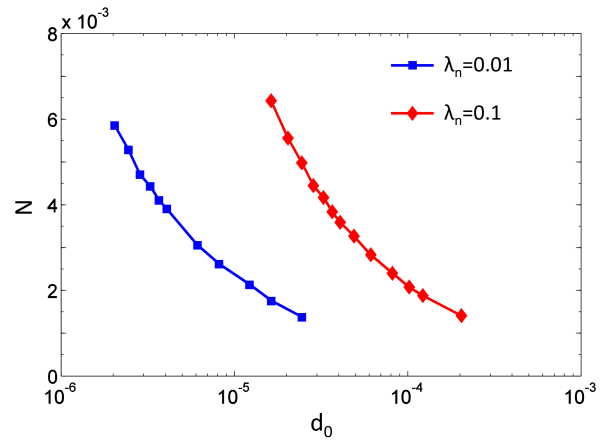}
\caption{The island density at a coverage of $\theta=0.1$ depends on both $d_0$ and $\lambda_n$. Simulations are done on a lattice with $1920\times1920$ grid points. $D/F=10^6$. }
\label{fig:capillary}
\end{figure}

We have pointed out (in connection with Fig~\ref{fig:side}) that nucleation is treated rather differently in the phase field model than in KMC or LSM simulations. To emphasize this point, Fig.~\ref{fig:capillary} shows the total island density as a function of  $d_0$ and $\lambda_n$ for $D/F=10^6$. The decrease in island density with increasing $d_0$ is striking, but not hard to understand. Larger $d_0$ increases the relative magnitude of the first two terms of the right hand side of Eq.~(\ref{phase}), which preserves the equilibrium state (i.e. $\phi=0$ or $\phi=1$). Consequently, proto-islands hardly grow in the beginning (when $\phi$ is close to zero)  and many of them diffuse away. The  island density increases as  $\lambda_n$ increases also. This parameter is the  coefficient of the nucleation term in (\ref{phase}). Given the same surrounding adatom concentration, as one adatom is deposited, a larger $\lambda_n$ triggers a larger change of the order parameter, which is more likely to survive and become an island.

\begin{figure}
\includegraphics[scale=0.45]{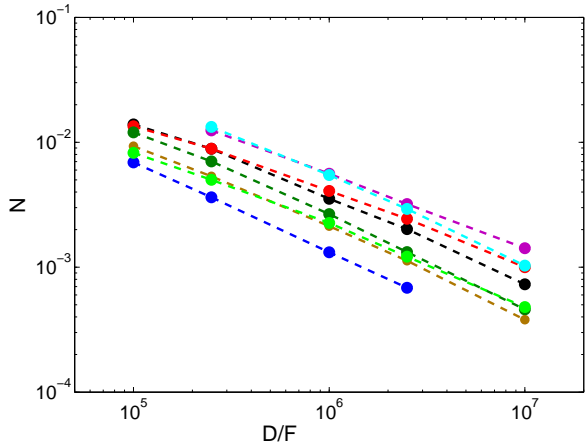}
\caption{The island density scaling vs. $D/F$ with different $d_0$ and $\lambda_n$. $d_0 = 1.44\times10^{-6} - 3.25\times10^{-4}, \lambda_n = 0.0084 - 1$, and $\theta=0.1$. }
\label{fig:scaling}
\end{figure}

The foregoing may be compared with a rate equation analysis or an LSM simulation, where the nucleation rate is determined by a {\it global} average of the adatom concentration over the whole domain. Specifically,
\begin{equation}
d N/d t = D\sigma_1 \langle u^2\rangle , \label{evol}
\end{equation}
where $\sigma_1$ is the (constant) capture number. In the standard rate theory of irreversible aggregations, (\ref{evol}) leads to a well-known scaling law for the total island density: $N \sim (D/F)^{-\chi}$ with $\chi=1/3$. This is also seen in irreversible LSM and KMC simulations. However, the mechanism implied by (\ref{evol}) is not truly captured by (\ref{density}) and (\ref{phase}). Instead, our phase field model uses $\lambda_n D u^2$ as a {\it local}  estimate of  the nucleation rate. We remind the reader that, unlike other simulation methods, most islands grow out of the initial adatom depositions in the phase field method. Be that as it may, upon fixing $d_0$ and $\lambda_n$ and changing only $D/F$, we found that the total island density shows distinct scaling behavior. This is shown in  Fig.~\ref{fig:scaling}. The curves of different color correspond to different values of $d_0$ and $\lambda_n$ over a wide range. The average value for the scaling exponent is $\chi \approx 0.65$. It is worth remarking that the island size distributions from different data points on the same curve in Fig.~\ref{fig:scaling} usually do not collapse very well. This suggests that the degree of reversibility is not the same. 

\begin{figure}
\includegraphics[scale=0.53]{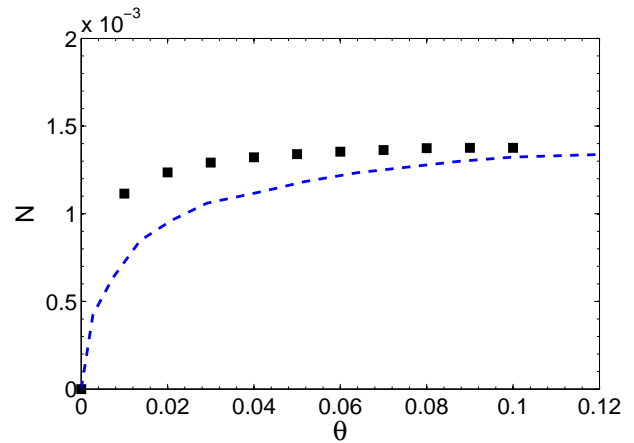}
\caption{Nucleation shuts off faster in phase field simulations. For $D/F=10^7$ and the same island density in the steady state, the time evolution of island density in phase field simulations (black squares) reaches the steady state much faster than in KMC simulations (dashed line, replotted from \cite{Ratsch2001}). $d_0=1.44\times10^{-6}$, $\lambda_n=8.4\times10^{-3}$. }
\label{fig:time}
\end{figure}

We do not fully understand the scaling seen in Fig~\ref{fig:scaling}, although we presume a simple analytic theory exists which can reproduce the observed exponent. On the other hand,  we can gain some insight by looking into the time evolution of the island density in more detail. Fig.~\ref{fig:time} is a typical curve of $N(t)$ obtained from a phase field simulation with $D/F = 10^7$. By changing the model parameters as described in Fig.~\ref{fig:capillary}, we can match the island density produced by a KMC simulation with the same value of $D/F$. However, there is a clear discrepancy in the nucleation rate: the island density approaches the steady state much faster in our simulations than in the KMC simulations. In fact, all of our phase field simulations show similar behavior. Since the island size distribution is a characteristic of the aggregation regime, this could explain why we can obtain the scaling of island size distribution at a much lower coverage than expected from  KMC simulations (see Fig.~\ref{fig:distr}). The fact that most islands tend to form at an earlier time is undoubtedly caused by the initial adatom depositions (see Fig.~\ref{fig:side}). It follows that the nucleation rate in this phase field model decreases faster than what we expect from Eq.~(\ref{evol}), which results in a stronger dependence on $u$ and thus changes the scaling of the island density.

\section{Conclusion}

In summary,  we have shown that phase field modeling of sub-monolayer epitaxial growth reproduces the scaled island size distributions seen in experiment and obtained from other high-quality simulation methods. The crossover from irreversible aggregation to reversible aggregation is driven by the magnitude of a capillary constant which enters the Gibbs-Thomson equation. This shows that diffusion-limited aggregation phenomena are well-captured by the model. \cite{Provatas1999} On the other hand, the scaling of the island density itself disagrees with experiment and with other simulation methods. This implies  that our model does not treat nucleation as accurately as one would like. One simple solution is to abandon the term $\lambda_n D u^2$ in (\ref{phase}) and use the level set method strategy to nucleate new islands. We suspect this will produce the correct total island density without changing the high quality already obtained for the island size distributions. This might be important, moving forward,  because the phase field method is less computationally intensive than the LSM and is much easier to implement at larger spatial scales and for  more complicated epitaxial growth situations.

\section{Ackowledgements}
We thank N. Goldenfeld and J. A. Dantzig for introducing us to the phase field method. We also thank J. W. Evans and C. Ratsch for helpful discussions. This research was supported in part by the National Science Foundation through TeraGrid resources provided by Texas Advanced Computing Center (TACC) under Grant No. TG-PHY100006. Fan Ming was supported by the MRSEC program of the National Science Foundation under Grant No. DMR-0820382.

\newpage
\bibliography{Phase_Field_Paper}% Produces the bibliography via BibTeX.

\newpage

\end{document}